# Integration studies of RF solid-state generators in the electrical system of NBTF experiments and ITER HNB

R. Casagrande[1], A. Maistrello[1], M. Recchia[1,2], M. De Nardi[1], M. Bigi[1], L. Zanotto[1], M. Boldrin[1], H. Decamps[3]

[1] *Consorzio RFX, Corso Stati Uniti 4, I-35127 Padova, Italy*
[2] *CNR – ISTP Padova, Italy*
[3] *ITER Organization, Route de Vinon-sur-Verdon, CS 90 046, 13067 St Paul Lez Durance Cedex – France*

SPIDER operation, started in 2018, pointed out performance-limiting issues caused by the technology employed in RF generators, based on tetrode free-running oscillators. One of these limits, namely the onset of frequency instabilities, prevented operation at the full rated power of 200 kW. In addition, tetrodes require high voltage to operate, which translates to risk of flashovers and the necessity to perform conditioning procedures, limiting the overall reliability. These disadvantages, combined with the positive experience gained in the meanwhile on smaller facilities with solid state amplifiers, led to the proposal of a complete re-design of the radiofrequency power supplies. This paper describes the modelling activities used to define specifications and design criteria of the solid-state amplifiers for SPIDER and MITICA, which can be directly transposed to the ITER HNB units when their functionality is proven. We detail the topology of the generators, consisting of class D amplifier modules combined to achieve the required 200 kW, which design is mainly driven by the necessity to deliver nominal power to the ion source, mitigate the risk of obsolescence, and improve the reliability through modularity. Due to the non-standard application, we gave particular focus to the integration of generators in the RF systems of SPIDER and MITICA. Numerical analyses were performed to verify the impact of harmonic distortion on transmission line and RF load components, to address the effect of mutual coupling between RF circuits on the generator output modulation, and to assess the magnitude of common mode currents in the electric system. These studies, as well as the experience gained from SPIDER operation, helped to define dedicated circuit design provisions and control strategies, which are currently being implemented in the detailed design and construction phase of the new RF amplifiers.

## I. INTRODUCTION

ITER Heating Neutral Beam injectors (HNB) [1] will exploit Radio-Frequency (RF) plasma drivers for the generation of negative ions of deuterium, which are subsequently extracted and accelerated to the nominal energy of 1 MeV, delivering up to 16.5 MW of heating power to the burning plasma. The ITER NBI prototypes, namely Megavolt ITER Injector and Concept Advancement (MITICA) and Source for the Production of Ions of Deuterium Extracted from RF plasma (SPIDER), are both under development at the Neutral Beam Test Facility (NBTF) [2] in Padova, Italy. The negative ion source [3], which features the same design in SPIDER, MITICA and ITER HNB units, is composed of 8 RF plasma drivers, followed by a system of grids for the filtering of electrons (Plasma Grid - PG), and extraction (Extraction Grid – EG) of negative ions, and is installed inside a vacuum vessel. The ion source body can be biased with respect to the PG, in order to reduce the electron density in front of the extraction apertures. After extraction, the voltage applied between EG and Grounded Grid (GG) accelerates the negative ions to nominal energy. In SPIDER, the acceleration is performed in a single -100 kV stage, whereas in MITICA and ITER HNB, the beam is accelerated in five stages, each rated for -200 kV, to reach 1 MeV energy.

The Ion Source and Extraction Power Supplies (ISEPS) [4], hosted in the High Voltage Deck (HVD) [5][6] - an insulated platform at the acceleration potential - power the ion source constitutive elements. The power supplies connect to the respective loads through a multiconductor transmission line. The production of plasma inside the ion source is achieved by means of the RF system (see Figure 1), which is composed of:

- RF generators: 4 RF free-running tetrode-based oscillators rated for 200 kW each, operating in the frequency range from 0.9 to 1.1 MHz, installed in the HVD;
- Transmission lines: 4 coaxial rigid lines, with length of approximately 40 m in SPIDER and 100 m in MITICA;
- RF load: 8 inductively coupled plasma drivers, hosted inside the vacuum vessel. Each generator powers 2 RF drivers, which are connected in series and represent one of the 4 sectors of the ion source. The sector impedance (series of $R_D$, $L_D$) is adapted to 50 Ω by means of a capacitive matching network ($C_P$ and $C_S$). The load is variable, with its impedance depending on the presence of plasma and on the ion source operating parameters (pressure, magnetic field, source biasing).

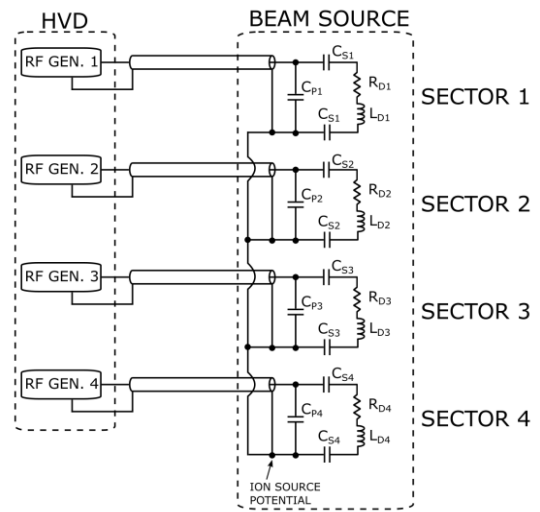

*Figure 1. RF system of NBTF experiments and ITER HNB, comprising RF generators, coaxial lines, matching networks and plasma drivers.*

The RF load is referred to the highest electrical potential of the system, i.e. the ion source potential. The sequence of power supplies lifting the ion source to its working potential are the acceleration power supply (AGPS), the extraction power supply (ISEG), the magnetic filter power supply (ISPG) and finally the source biasing power supply (ISBI). The layout of SPIDER electrical system is represented, with a simplified view, in Figure 2. For MITICA and ITER HNB, the acceleration system – in particular the transmission line [6] and AGPS [7] – features a higher degree of complexity with respect to SPIDER. However, the conceptual representation of Figure 2 is sufficient for both NBTF experiments and ITER HNB, as a reference for the topics discussed in this manuscript.

Since 2018, the operation of SPIDER has been focused on a steady increase of its performance, with the aim of reaching ITER requirements on the negative ion source. The achievement of nominal performance was prevented by several issues, mostly the occurrence of RF breakdowns in the beam source [8][9], limitations on the acceleration energy [10] and the problem to deliver the full 800 kW power to the RF drivers. In particular, the latter was mainly caused by a constraint of the free-running tetrode-based oscillators used as RF generators in SPIDER, namely the presence of a frequency instability which prevents the achievement of a 50 Ω matching [11]. In addition, oscillators feature high voltage vacuum components, which require careful maintenance and often suffer from electrical breakdown during the operation, with the subsequent necessity for high voltage conditioning. To overcome the issues intrinsically related to free-running oscillators, the RF generators of SPIDER and MITICA will be replaced with amplifiers based on solid-state technology, that are briefly introduced in Section II; ITER baseline was modified as well, considering solid-state amplifiers as power supplies for the RF ion source of its HNB units [12]. The new solid-state amplifiers are currently in the detailed design phase, with the first prototype to be completed in 2023.

Other experimental facilities already provided a positive feedback from the operation with broadcasting solid-state amplifiers, which have been adapted for the usage on RF ion sources [13]. Nonetheless, the complexity and the strict constraints of NBTF experiments and ITER HNB require to carry out some specific considerations for the integration of the new RF solid-state generators in the existing electrical system.

Some of these considerations are related to the peculiarities of the RF load of the ion source: its variability, and the experimental needs of SPIDER and MITICA, require the RF amplifier to work with a large degree of flexibility. In Section III we describe the specifications for the control and protection system of the new RF amplifiers, aimed for the operation in a wide range of working conditions.

Another aspect to consider in the integration process is the interplay of the RF amplifier specific design characteristics with the rest of the RF system: in Section IV, the issue of the amplifier output harmonic distortion is introduced, detailing how the specification on its maximum allowable value was defined.

In conclusion, the experience gained during previous SPIDER experimentation was applied to address some integration issues already observed in the past operation with RF oscillators: in Section V we describe the specifications on the power distribution system and reference potential connections for the new RF generators, which are aimed at mitigating the RF stray currents circulation and the transient overvoltage caused by breakdowns between the ion source grids.

## II. RFSSA FUNDAMENTAL STRUCTURE

The new RF Solid-State Amplifiers (RFSSA), designed with a modular structure, will be able to deliver in continuous-wave mode a nominal power of 200 kW to a 50 Ω load, in the frequency range from 0.92 to 1.08 MHz. The RFSSA units, represented schematically in Figure 3, feature the main following components:

- Input three-phase rectifier and DC power supply, to feed the DC link of the RF power modules.
- RF power modules, consisting of a MOSFET H-bridge, which produces an output square wave (class D amplifier). Unlike the free-running oscillators, the new RFSSA do not feature a self-oscillating circuit to establish the operating frequency.
- The RF modules output power is summed through a ferrite series combiner: each power module output is wound around a ferrite toroid, constituting the primary of a transformer. The secondary is a copper bar passing through the ferrite toroid central aperture, fundamentally representing a single winding. One end of the bar is connected to the RFSSA cubicle reference potential, while the other end represents the RF power output. In this configuration, the RF voltage from the power modules sums along the length of the combiner secondary.
- The impedance needed at the combiner output for maximum power transfer from the RF power modules is not 50 Ω, which are instead required for the coaxial line connecting the RFSSA to its load. Therefore, a matching network is installed between the secondary of the combiner and the RFSSA coaxial output. The matching network serves also as a low-pass filter, to obtain an output sine wave starting from the RF power modules square waves.

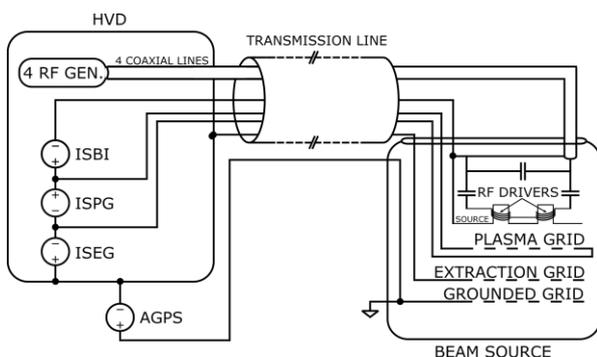

*Figure 2. Simplified representation of SPIDER electrical system layout. For simplicity, the medium and low voltage power distribution is not represented.*

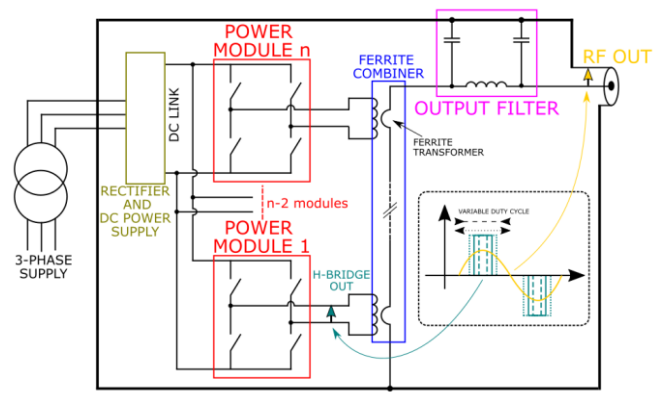

*Figure 3. RFSSA unit with its fundamental blocks: DC supply, RF power modules, power combiner, output filter and matching network. The power module and RFSSA output voltages are represented in green and yellow, respectively.*

## III. RFSSA CONTROL AND PROTECTION SPECIFICATIONS

The requirements on the RFSSA control system have been derived from the operational experience with SPIDER: the main objective is to simplify the tasks to be performed by the operator, when managing the RF system, in order to avoid as much as possible human errors and unwanted shutdowns of the amplifiers. For the scientific exploitation of SPIDER, a common technique used to study the machine performance is the parametric scanning, which can significantly impact on the RF load impedance. The control system specifications are based on the philosophy that the RFSSA unit shall be able to work in a wide range of conditions, adapting automatically to the load and allowing a continuous operation without unwanted shut-downs induced by transient events such as plasma initiation, plasma loss and breakdown in the RF circuits onboard the ion source.

### A. Power control

#### 1) VSWR feedback

The RF modules output power, and consequently the total RFSSA power, is varied through the adjustment of the duty cycle of the square wave produced by the H-bridges. The RFSSA output power is monitored through a dual directional coupler, from which the information of the total power delivered to the load is extracted and fed to the control loop, allowing the automatic setting of the required duty cycle.

The RFSSA can operate at full rated power only in a neighborhood of Voltage Standing Wave Ratio (VSWR) equal to 1. Outside of this range, the amplifier is subject to a power de-rating. To avoid a shutdown of the generator when it experiences excessive mismatching, an automatic system limits the available output power according to the VSWR obtained in real time from the dual directional coupler measurements. The control system carries out another limitation, taking into account that the driver equivalent resistance is low prior to plasma initiation (vacuum load), so excessively large voltage can develop across the coil or the matching network capacitors if high RF power is delivered to the load. For this reason, the RFSSA output power is limited to $P_{VL}$ for values of VSWR larger than the minimum foreseen for the vacuum load ($VSWR_{VL}$), as represented in Figure 4.

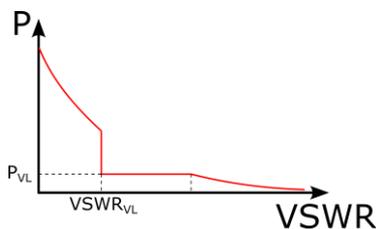

*Figure 4. Automatic power limitation, depending on VSWR measured value.*

#### 2) Power burst

It is a common occurrence to observe sparks on high voltage components of the RF system upon restart of SPIDER operation, after shutdowns in which the vacuum vessel was opened. Typically, several conditioning shots are necessary to re-establish the nominal working conditions of the RF elements installed inside the vacuum vessel, namely the sections of coaxial line, the matching network capacitors and the plasma drivers' coils. To simplify this procedure, and in general to avoid the termination of an experimental shot in case of a single spark event, the RFSSA was designed to perform the burst sequence: when the VSWR exceeds a specified limit, the generator shuts down and rapidly turns on again, gradually delivering back the requested power to the load. A delay between bursts can be set by the operator, thus the total time needed to return to operational conditions depends on this delay and on the requested power, considering that the specified power derivative for the RFSSA is 10 kW/ms. After a predefined number of burst sequences has been performed, with the VSWR not going back to nominal values, the RFSSA shuts down and the control system raises an error, signaling the probable presence of a fault in the RF circuit.

### B. Frequency control

Since the matching network installed in the ion source employs fixed capacitors, the minimum power reflection conditions, or equivalently minimum VSWR, can be achieved solely via the adjustment of the operating frequency. To simplify this procedure, the matching is performed through an automatic algorithm, which monitors in real time the VSWR at the RFSSA output, and looks for its minimum by changing the output frequency.

The frequency control can be performed in two modes, depending whether it is required to operate the RFSSA units independently, or in a synchronized way:

- Independent frequencies mode: independent control of each RFSSA unit, either manually or via the automatic matching algorithm, in order to achieve the best possible matching for all generators.
- Iso-frequency mode: operation of all RFSSA units at the same frequency. In this case one generator acts as the master, with manual frequency setting or control through the automatic matching algorithm, and the other units follow the master reference.

The type of frequency control can affect the operation of RFSSA units, due to the RF load peculiarities: the physical routing of connections from matching network capacitors to RF drivers coils causes a non-negligible mutual coupling between the four RF circuits inside the beam source. This coupling introduces a variation of the impedance seen by one RF generator, when one or more of the other amplifiers is operating, depending on the operational mode. In SPIDER, we observed that highest mutual coupling is between sectors 1 and 2 of the beam source, and consequently for sectors 3 and 4, which feature the same routing of the connections. In all the other combinations of sectors the mutual coupling is negligible, so the considerations carried out in the following paragraphs can reduce to two RF generators at a time.

#### 1) Independent frequencies mode

In this operational mode, each generator is controlled independently, with the matching algorithm searching for the frequencies to achieve minimum VSWR for all RFSSA units. When operating two generators with different frequencies $f_1$ and $f_2$, the mutual coupling causes an amplitude modulation of the voltage and current at the RFSSA output, at a frequency corresponding to the difference $|f_1-f_2|$. This modulation has a negative impact on the RFSSA operation, because it can cause the output voltage or current to exceed the limits, when multiple generators work simultaneously at full-rated power. In addition, the modulation translates into a variation in time of the VSWR, which can cause instabilities of the automatic matching algorithm. To reduce these problems, the operating frequency of the generators needs to be chosen such as to minimize the interaction between RF circuits. This is achieved by covering as much as possible the available spectrum (from 0.92 to 1.08 MHz), in order to operate circuits with high mutual coupling at frequencies far apart from each other. Practically, this means that the matching networks need to be modified to be able to operate the circuits at different frequencies and with minimum VSWR, since in the present SPIDER configuration all the matching capacitors and RF drivers have the same impedance, within the tolerances. As an example, let us assume that the matching networks of two RF loads are tuned to achieve the minimum VSWR at the desired frequencies $f_{1-OPT}$ and $f_{2-OPT}$, without considering the mutual coupling between them. If the two generators are operated simultaneously at the same power, the VSWR profile is going to be subject to a modulation in time, potentially reaching the RFSSA limit at the highest peak. This is represented in Figure 5, where a MATLAB Simulink circuital model representing two coupled RF circuits was used to compute the VSWR modulation for different values of $\Delta f = f_{1-OPT} - f_{2-OPT}$, with a mutual coupling that represents the highest value measured in SPIDER, namely M = 0.05 µH. The deviation from the ideal VSWR decreases with $\Delta f$: in Figure 5a we see that, with $\Delta f$ = 25 kHz, at the peak of modulation the VSWR at the matching frequency is 11% larger than the VSWR achievable without coupling between circuits. By increasing $\Delta f$ to 50 kHz (Figure 5b) and to 75 kHZ (Figure 5c), the discrepancy between modulated VSWR and VSWR with no coupling reduces respectively to 9% and 7%.

Nonetheless, the modulation is still present even when working at the edges of the RFSSA available spectrum. Therefore, this method guarantees a mitigation of the VSWR modulation, but it cannot remove it completely. To decrease the VSWR modulation to negligible values, it would be necessary to further reduce the mutual coupling between RF circuits onboard the ion source, considering that for a given $\Delta f$, the amplitude of the modulation depends linearly on the mutual inductance.

*2) Iso-frequency mode*

In this operational mode, all the generators work at the same frequency, with the matching algorithm searching for the optimum VSWR referring only to the measurements from the master unit. In this case, there is no modulation in time of the voltage and current at the output of the generators, when operating more RFSSA units simultaneously. Nonetheless, if there is a non-negligible mutual coupling between the RF circuits, the impedance seen by two generators working at the same frequency and same power on two identical RF loads (same matching capacitors and same RF drivers' impedances), varies depending on the phase shift Φ between voltages at the matching network input. The load changes in opposite directions for the two generators, i.e. one RFSSA unit sees a higher impedance and the other a lower impedance with respect to the ideal case with no coupling, apart from the conditions in which Φ = 0° and Φ = 180°, where both RFSSA units see the same load. From the point of view of the VSWR, this means that the optimal conditions can be reached simultaneously for two RFSSA units working at the same frequency, only when Φ is 0° or 180° (see Figure 6). For all the other phase shift values, one of the two generators will see a higher VSWR with respect to the case in which it is operated singularly, thus it might not be able to deliver the nominal power when a second generator is switched on. In the specific case reported in Figure 6, with M = 0.05 µH, the VSWR at the matching frequency can be up to 9% larger than the VSWR without coupling between circuits, if the phase is not controlled.

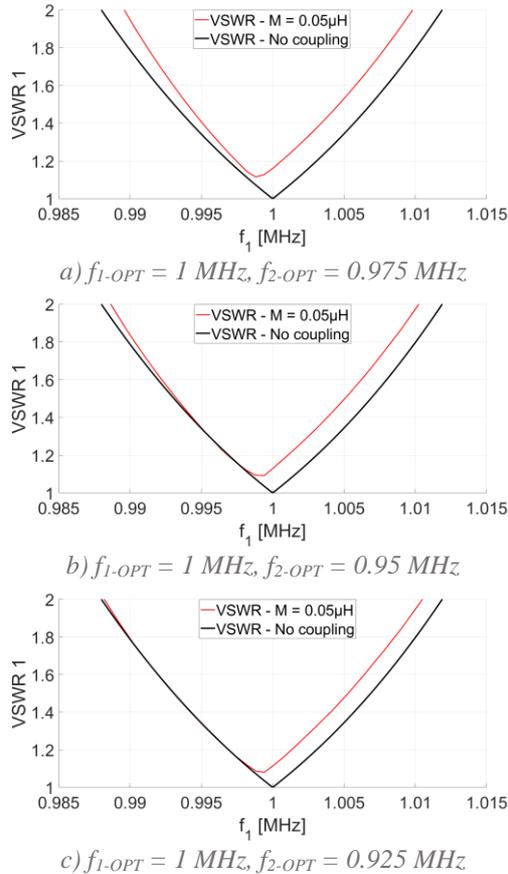

*a) $f_{1-OPT}$ = 1 MHz, $f_{2-OPT}$ = 0.975 MHz*

*b) $f_{1-OPT}$ = 1 MHz, $f_{2-OPT}$ = 0.95 MHz*

*c) $f_{1-OPT}$ = 1 MHz, $f_{2-OPT}$ = 0.925 MHz*

Figure 5. VSWR modulation when two generators supply mutually coupled RF loads. The red trace represents the peak value of the modulation, while the black trace is the VSWR profile in the ideal case, without coupling between circuits.

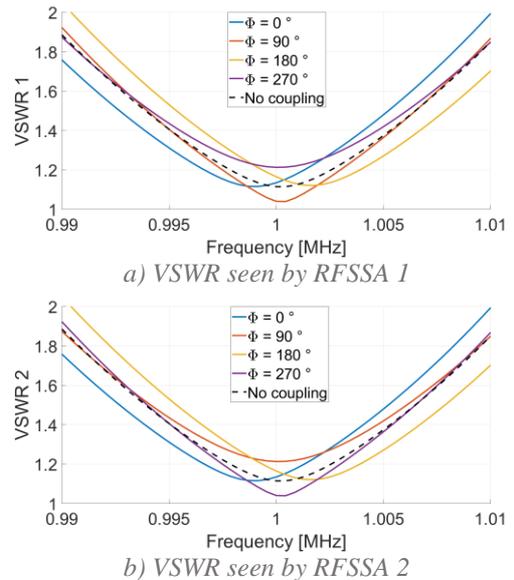

*a) VSWR seen by RFSSA 1*

*b) VSWR seen by RFSSA 2*

Figure 6. VSWR variation, depending on phase shift Φ between voltages at the input of the matching network. The black dashed line represents the ideal case, with no mutual coupling between RF circuits.

## C. Phase control

When operating the solid-state amplifiers in iso-frequency mode, an additional parameter that defines the status of the system can be introduced, namely the phase shift between RFSSA units output voltage. As described in the previous paragraph, the phase-shift has an effect on the simultaneous operation of multiple RF amplifiers, potentially preventing the achievement of nominal power. To avoid this issue, assuming that the impedance of different sectors can be matched at 50 Ω at the same frequency, the phase between voltages at the input of the matching networks shall be set to specific values ($\Phi = 0°$ or $\Phi = 180°$). Therefore, in the RFSSA units it will be possible to control the phase of the MOSFETs gate commands, in order to adjust the phase shift between voltages at the output of the generators. For the initial operation of the RFSSAs, we did not foresee a closed loop control on the phase shift; there is nonetheless the possibility to interface with an external control system capable of performing this action, if the first tests in SPIDER point out the necessity to automatically regulate the phase-shift to operate at nominal power.

## IV. SPECIFICATIONS ON HARMONIC DISTORTION

Differently than RF oscillators, the RFSSA power modules will not directly produce a sine wave at their outputs, but a square wave with variable duty cycle, which contains odd harmonics, multiple of the fundamental operating frequency. For specific lengths of the transmission line, the system can become resonant at the frequencies of higher harmonics, exceeding the output current limit of the RFSSA – leading to its shutdown – or causing overvoltage on the RF system components. Therefore, we need to define what is the maximum acceptable Total Harmonic Distortion (THD) of the RFSSA output voltage, i.e. how to size the low-pass filters in order to avoid damage of components.

Figure 7 shows the impedance seen by the generator, considering the range of SPIDER (40 m) and MITICA (100 m) transmission line lengths. The MITICA-like transmission line length is representative also of ITER HNB units.

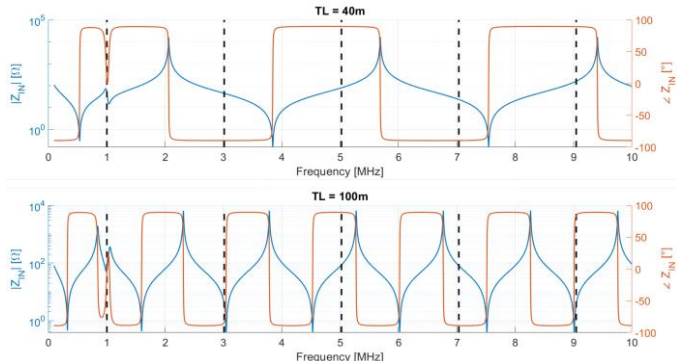

*Figure 7. Impedance seen by the RFSSA, for SPIDER and MITICA typical transmission line lengths. The black dashed lines highlight the fundamental harmonic, and the first four odd harmonics (3rd, 5th, 7th and 9th) produced by the RFSSA.*

We observe that, in the case of a 100 m transmission line, the third harmonic of the operating frequency is close to a resonant point, corresponding to a low impedance. In such a configuration, three operational problems arise:

- A standing wave develops, creating voltage anti-nodes where the dielectric strength of the coaxial line can be exceeded.
- High current is drawn from the RFSSA, potentially causing its shutdown.
- The matching network capacitor $C_P$ acts as a filter for the higher harmonics, thus a large current flows through it in case of resonance.

To analyze how much the harmonic content at the RFSSA output need to be reduced to avoid the aforementioned issues, we computed the ratio $\frac{I_n}{I_1}$ between a specific current harmonic and the fundamental, with $n$ harmonic number, considering SPIDER and MITICA relevant transmission line lengths, and several load conditions. In particular, we have taken into consideration the following parameters in the analysis:

- Transmission line lengths in the range from 35 to 45 m and from 100 to 120 m.
- Third and fifth harmonic of the operation frequency, defined considering the working point with minimum VSWR.
- Nominal plasma load, with $R_D = 4.5$ Ω; $L_D = 19$ μH [14].
- Fixed matching network with $C_P = 10$ nF, $C_S = 3$ nF.

The third and fifth harmonics amplitudes have been calculated assuming a square wave at the RFSSA output, thus considering that the generator is delivering the maximum power (100 % duty cycle). For each working frequency, the transmission line length for which there is the maximum current ratio $\frac{I_n}{I_1}$, with n=3, 5 was identified. The worst-case conditions were identified for MITICA-relevant transmission line lengths, and for the third harmonic. The blue trace in Figure 8 represents the ratio between the third harmonic of current and the fundamental, in the operating frequency range.

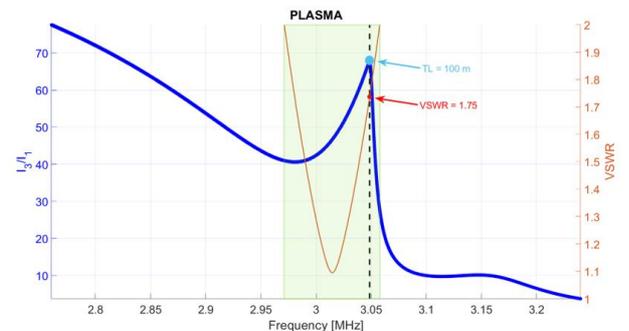

*Figure 8. Increase of the third harmonic of current drawn from the RFSSA unit, due to resonances caused by specific transmission line lengths.*

Each point corresponds to a specific transmission line length, for which the ratio $\frac{I_3}{I_1}$ is maximum. Based on the information available at the time of the analyses, and taking into account a sufficient margin, we consider that the RFSSA is capable to deliver the nominal power up to VSWR = 2. Therefore, the available operating frequency range reduces to the green area highlighted in Figure 8. Inside this area, we can identify a combination of transmission line length and frequency that causes the most unfavorable conditions for the third harmonic amplification: this set of parameters is used as the case study to identify the maximum allowable THD of the RFSSA output voltage. For this study, we considered a low-pass filter installed at the generator output, sized to achieve a specific set of desired THD values from the filtering of a square wave. From this set, the amplitude of the residual third harmonic was then extrapolated, and used to compute the resulting maximum voltage along the coaxial transmission line, and the voltage and current on the matching capacitor $C_P$. The results obtained are resumed in Table 1, where the percentage of the aforementioned quantities with respect to the components limits are shown for different values of THD. In case of resonance, the third harmonic of voltage and current in the transmission line and matching network are within the components limits for a THD below 1%, so this specification was provided to the supplier to define the sizing of the low-pass filters in the power modules and at the output of the RFSSA. It is highlighted again that these computations refer to the worst case scenario, i.e. the situation in which the third harmonic of the operating frequency resonates at a specific transmission line length. Even though the voltage and current on critical components are within the limits, considering the sum of third harmonic and fundamental, it would be preferable to avoid altogether the worst case operating condition. Therefore, if resonances are encountered during operation in steady state, they can be avoided by adding a short section of coaxial transmission line inside the HVD.

Keeping the THD to very low values is nonetheless fundamental to avoid damage on components or shutdown of the amplifier, since additional harmonics not considered in this analysis could be excited during transients (e.g. plasma initiation, plasma shutdown or fast power adjustments). To validate the measures taken to limit the THD in the design of the solid-state amplifiers, dedicated tests will be carried out on the prototype RFSSA unit, using a resonant dummy load with a variable-length transmission line [15] to reproduce the configuration that causes a resonance for the third harmonic. Several operational sequences will be tested on the resonant dummy load, in order to reproduce as closely as possible the transients during plasma initiation, plasma shutdown and power adjustment. Modifications to the overall RFSSA output filter are relatively simple to introduce, therefore if issues are observed during the dedicated tests on the dummy load, the filter design could be reviewed to further reduce the THD, with a limited impact on the schedule.

|  | NO FILTER | THD 10% | THD 5% | THD 1% |
|---|---|---|---|---|
| $I_{CP}/I_{CP\text{-}MAX}$ [%] | 425 | 132 | 70 | 34 |
| $V_{CP}/V_{CP\text{-}MAX}$ [%] | 446 | 188 | 131 | 11 |
| $V_{TL}/V_{TL\text{-}MAX}$ [%] | 1700 | 516 | 258 | 50 |

Table 1. Percentage of third harmonic of voltage and current on matching capacitor $C_P$ ($V_{CP}$, $I_{CP}$) and voltage on Transmission Line ($V_{TL}$) with respect to their operating limits.

## V. POWER DISTRIBUTION SYSTEM AND POTENTIAL REFERENCES OF RFSSA UNITS

The AC power distribution system of the new RFSSA units differs significantly from the previous one used for the oscillators, mainly for the following reasons:

- Characteristics of the AC system load: in the configuration with oscillators, the main load of the AC distribution system is a high voltage DC power supply (12 kV, 128 A), which is used to feed the tetrode anode. For the new RFSSA units, the main load of the AC system is instead a three-phase rectifier, which supplies a voltage regulator and, following, the DC-link of the H-bridges. This AC load requires low voltage and high current.

- Insulation from HVD: the tetrode-based oscillators feature an output transformer that serves multiple purposes, namely matching from the optimal impedance required by tetrodes to the 50 Ω of the transmission line, and insulation between the RF output reference potential (ion source potential) and the rest of the oscillator components, referred to the HVD. The new RF generators will not have an output RF transformer, meaning that the coaxial lines directly refer the cubicles to the ion source potential. Therefore, the RFSSA units need to be insulated from the HVD by means of fibre-reinforced plastic beams.

The main implication of these aspects is that the RFSSA units will be supplied through isolation transformers, since the medium voltage distribution in ISEPS is referred to the HVD potential. In addition, to reduce the harmonic distortion of the total current absorbed by the three-phase rectifiers installed in the four RFSSA units, the transformers will be configured so to resemble the current absorption of a 12-pulse rectifier. This is achieved using transformers with vector group Yy0 to supply two RFSSA units, and with vector group Dy11 to supply the other two units. The resulting fundamental scheme is represented in Figure 9.

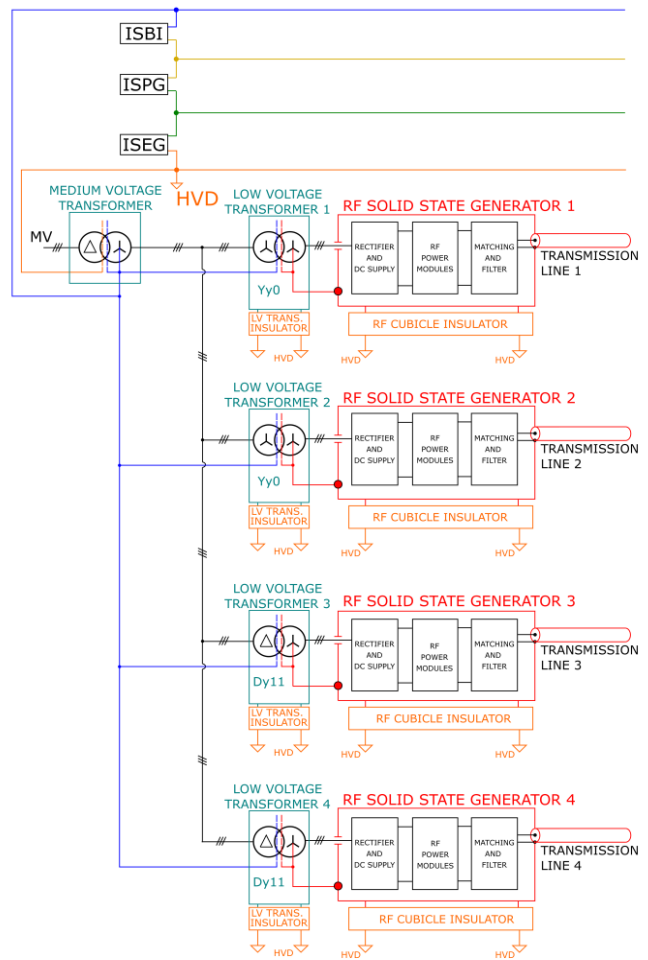

Figure 9. Reference power distribution scheme and potential reference connections for the new RFSSA units.

The electrical issues that emerged from early SPIDER operation, motivate the necessity for dedicated studies on the integration of RFSSA units in the complex electrical systems of SPIDER, MITICA and ITER HNB, in particular with respect to:

- RF stray currents: operation of SPIDER at high RF power has revealed the presence of large RF stray currents circulating in non-RF power supplies and relative components in ISEPS [16][17], causing damage due to overheating and noise on several diagnostics.
- High voltage breakdown: the power supplies for the extraction and acceleration stages of NBTF experiments and ITER HNB, are designed to operate in conditions of repetitive electrical breakdown between grids. A major issue that characterized the operation of SPIDER at large acceleration energies was the presence of high voltage transients following the breakdown, which either triggered spurious faults or damaged the sensitive components of the power supplies control boards.

These issues are both dependent on the potential reference connections of the RF system, as well as the layout of its AC power distribution.

### A. Mitigation of RF stray currents

The issue of RF stray currents circulation has been extensively studied in the configuration of SPIDER ISEPS with RF oscillators [8][16][17][18], leading to the development of a solution to mitigate them, based on the re-routing of the oscillators capacitive connections towards the HVD [19]. However, the main consideration carried out for the oscillators is not applicable to the new RF system with solid-state amplifiers, since the driving phenomenon and the circuital scheme to consider is significantly different. Therefore, to study the integration of RFSSA units power distribution into ISEPS, we initially focused our considerations on how to minimize the RF stray currents circulation in this new system. These currents are fundamentally generated by the longitudinal voltage drop along the outer conductor of the RF coaxial line. The two ends of the coaxial line are connected together via several other conductors (ion source bias conductor, PG conductor) and capacitive elements, thus forming a closed loop and allowing the circulation of a common mode current through the components of ISEPS. One of the capacitive connections that affect the circulation of stray currents in the new RF system, depends on the configuration of the isolation transformers supplying the RFSSA units. The medium voltage and low voltage transformers are both equipped with electrostatic screens (see Figure 9), thus introducing a non-negligible stray capacitance between the reference potential of the primary winding and that of the secondary winding. The screens are connected as following:

- Medium voltage transformer:
    I. Primary screen referred to HVD
    II. Secondary screen referred to ISBI (ion source potential)
- Low voltage transformers:
    I. Primary screen referred to ISBI (ion source potential)
    II. Secondary screen referred to RFSSA cubicle (ion source potential)

Figure 10 shows that the most direct path for the stray current is through the stray capacitance between primary and secondary screens of the low voltage transformers. It is thus important to minimize this parameter to mitigate the circulation of stray common mode currents; this is achieved by increasing the distance between screens, so also between primary and secondary winding, with the downside of a larger and heavier transformer. Since the upgrade to RFSSA is happening after the final design and construction of SPIDER and MITICA HVDs, there are several constraints on the space and weight to be installed in them. This limits the available size for the low voltage transformers, hence the minimum achievable capacitance between screens. If this stray capacitance is not sufficiently low to maintain stray currents below acceptable values, we have foreseen the possibility to install an inductor in parallel to the low voltage transformers screens. If the inductance is tuned to resonate with the screens capacitance at the RFSSA working frequency (~ 1 MHz), the L-C parallel will constitute a high impedance for the common mode stray current, limiting its circulation. The inductance fundamentally represents a short circuit at low frequencies, but this connection is possible since the primary and secondary screens of the low voltage transformers are already at the same potential (ion source potential); the insulation from HVD is still guaranteed by the medium voltage transformer.

The alternative circulation path for stray currents is through ISEG filter and the capacitance of the RFSSA cubicle with respect to HVD. In this case it is not possible to install a filter to block the common mode currents, but their magnitude should in any case be much lower with respect to the ones measured in SPIDER with RF oscillators [19], for the following reasons:

- The capacitance between RFSSA cubicle and HVD can be kept to very low values, since the HVD surface facing the cubicle is actually a metal grating.
- The common mode driving voltage in this configuration depends on the voltage drop along the RF coaxial line. This has a much lower amplitude with respect to the configuration with RF oscillators, where the stray currents were driven by the voltage difference between primary and secondary winding of the RF output transformer.

With these provisions, it should be possible to maintain the RF stray currents to negligible values in NBTF experiments and ITER HNB electrical systems.

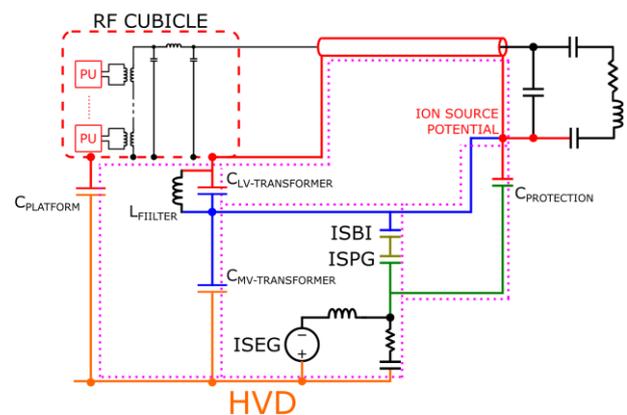

*Figure 10. RF stray currents circulation paths (in magenta) in the new RF system configuration. The most direct path is through the conductor biasing the ion source (in blue) and the low-voltage transformer screens capacitance. The optional filter inductance in parallel to this capacitance is also represented.*

## B. Immunity to breakdown between grids

In the beam source of NBTF experiments and ITER HNB, the breakdown in the acceleration and extraction stages is a common occurrence during experimentation, and it has been considered as a nominal operating condition for the design of the power supply system. For this reason, we also need to verify if the modification introduced to the RF system is compliant with this aspect, i.e. if the new configuration is immune to the breakdown transients.

To describe the problems related to the breakdown with the simplest possible picture, we can consider a capacitance charged at the acceleration voltage $V_{ACC}$, which discharges through an inductance because of an arc inside the beam source (see Figure 11).

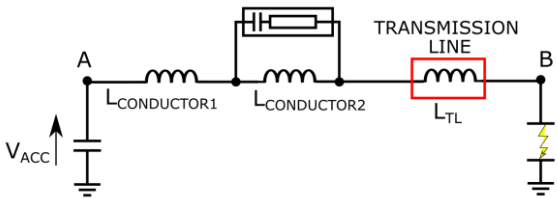

*Figure 11. Simplified representation of the circuit causing oscillatory transients at the breakdown between grids.*

The capacitance represents the actual HVD, the AGPS output stage, and all the components at the acceleration voltage and insulated from ground, whereas the inductance represents the connections between HVD and beam source ($L_{CONDUCTOR1}$, $L_{CONDUCTOR2}$, and $L_{TL}$).

When a breakdown happens, the capacitances and inductances in this system give rise to an oscillatory transient, in the hundreds of kHz to MHz range [20]

In the real picture, the situation is much more complex (see Figure 12):

- When the capacitances at acceleration voltage discharge, there are several possible paths for the breakdown current. From the HVD, the current can flow through the MV transformer or through the DC power supplies output stages, then closing the path via the bias, PG, EG or coaxial line conductors.
- The electrical system comprises a large number of equivalent reactive elements (mainly stray) participating in the breakdowns transient, which cause oscillations in a broad spectrum of frequencies.
- During transients, the elements that represent a high impedance at the characteristic frequency of the oscillations can experience high voltage, up to the acceleration potential.

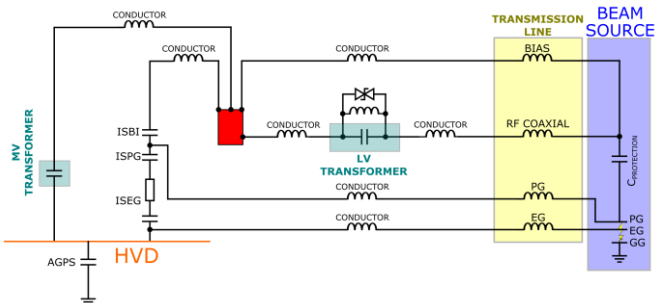

*Figure 12. Highlight of the possible breakdown current paths which involve the power distribution system for the new RFSSA.*

Issues related to the breakdown transients were observed in SPIDER, during operation at acceleration voltage higher than 30 kV, in the control system of power supplies at the ion source potential, thus insulated from the HVD. Unlike the RF system with oscillators, the new solid-state generators are also insulated from the HVD, and supplied by isolation transformers. Therefore, the experience gained from the tests and analyses carried out to solve SPIDER breakdown immunity problems, was used to identify potentially critical issues related to the new RF system configuration.

### 1) Low voltage transformers

To reduce the circulation of RF stray currents, the low voltage transformers need to have a low stray capacitance between primary and secondary electrostatic screens. There is an alternative option to block these current, namely the installation of an inductor $L_{FILTER}$ in parallel to the screens, with the aim of achieving a resonant LC filter for the common mode current. Regardless of which final configuration will be chosen, i.e. only screen capacitance or resonant filter, in the range of frequencies considered (hundreds of kHz to MHz), the transformers will represent a high impedance. Since these components are in the breakdown path, is necessary to protect the transformers during transients, with a voltage suppressor or crowbar to prevent arcs between the screens or windings. Integrating this protection in the system is not a critical aspect, since the potential of the primary and secondary screens is the same, thus the transient suppressing device does not need to sustain a high voltage during normal operation. The device shall nonetheless have a small capacitance, for the case in which the resonant filter is not installed, to maintain a high impedance for the RF stray currents.

### 2) Medium voltage transformer

The medium voltage transformer also features a small capacitance between electrostatic screens, thus during the breakdown transients it can experience voltage peaks higher than its holding capabilities. Since the medium voltage transformer is also used to guarantee the insulation of the RFSSA units, the transient voltage suppressor system needs to sustain the extraction voltage applied by ISEG (up to 12 kV) during normal operation.

### 3) Electronics of the control system

One of the major issues experienced during breakdown between grids in SPIDER was the occurrence of spurious faults, triggered by the control and protection system of power supplies at the ion source potential. These faults were not caused by real overvoltage or overcurrent at the power supplies output, but were triggered because the electronic boards of the control system experience transient overvoltage at the breakdown, producing spurious spikes on their output signals that are interpreted as faults by the central control unit. The cause of this transient overvoltage can be described referring to Figure 11: the capacitances at acceleration voltage are decoupled by the transmission line connections, therefore they discharge at different rates, depending on their distance from the breakdown point, i.e. from the grids. This causes a voltage drop between points A and B, thus if sensitive equipment is referred to multiple points along this path, e.g. the components in parallel to $L_{CONDUCTOR2}$ in Figure 13, it might experience significant overvoltage at breakdowns.

Applying such transient voltages on the sensitive electronic equipment of the control system, leads to the generation of spurious output signals and consequent triggering of faults, compromising the experimental session.

To avoid this occurrence on the RFSSA units, the system of transducers, electronic boards and supplies used for the control are installed inside the generator and connected to the reference potential in locations physically close to each other, so to minimize the inductance $L_{CUBICLE}$ (see Figure 13).

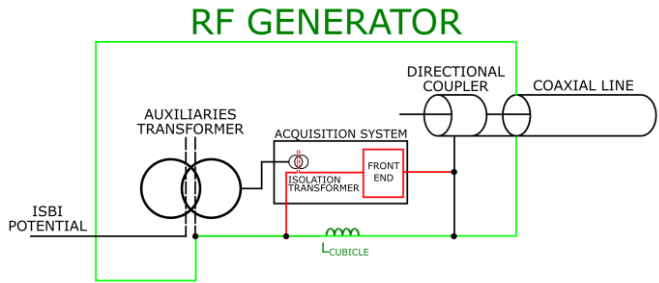

*Figure 13. Potential references of the measurement system used for the control of the RFSSA units.*

## VI. CONCLUSIONS

RF generators based on solid-state amplifiers are now being developed for the NBTF experiments and ITER HNB, to overcome the technological limitations of the presently installed tetrode-based oscillators. The design of these RF solid-state amplifiers is aimed at improving the reliability, availability and performance with respect to RF oscillators, due to the absence of high voltage components, a modular structure and better dynamics and precision of power and frequency control. Despite these advantages, the peculiarity of the RF system, the need for a significant experimental flexibility and the interplay between the installed power supplies and components onboard the ion source, require specific attention for the integration and operation of the new RFSSA units in NBTF facilities.

Due to the complexity of the ITER HNB system, it is of paramount importance to keep in mind the concepts of simplicity of operation and flexibility during the development of the RF amplifiers control scheme: we took advantage of the new technology to introduce automatic power handling and frequency matching systems, both based on VSWR measurements feedback, that should significantly reduce the number of tasks to be performed by the ion source operator with respect to the present system with RF oscillators.

In the past experimentation with SPIDER, an aspect that significantly downgraded the capabilities of oscillators was the spurious interaction between RF generators and RF load. For the RFSSA, we tried instead to consider in detail the possible interactions already in the design phase, introducing multiple frequency control strategies and limiting the output total harmonic distortion. The control stability, as well as the suppression of deleterious interactions with the RF load, are fundamental for the development of new effective RF generators: these aspect will thus be extensively tested during factory acceptance on a dedicated resonant dummy load, and during the site acceptance through the integration with the actual ion source.

Besides the considerations concerning the design of the RF amplifiers and the interaction with the RF load, we carried out an analysis on the AC power distribution layout, an aspect that is usually overlooked in the design of experimental power supplies. The experience from SPIDER suggests that, even though standard equipment such as transformers and transducers are functional in nominal conditions, they may be significantly affected by transients caused by the breakdown between grids or circulation of RF stray currents. The preliminary analysis presented in this paper was aimed at defining the general AC distribution layout, with specific provisions on the transformers configuration and potential reference connections, focused on the mitigation of common mode currents and immunity from breakdown. Detailed analyses are currently being carried out, in order to identify dedicated transient protections for the transformers and additional filters for RF stray currents. The provisions against RF stray currents will be tested during factory acceptance [15], whereas the breakdown immunity will be assessed after the installation of the RFSSA units in SPIDER ISEPS, following the same methodology described in [10].

## ACKNOWLEDGMENTS

This work has been carried out within the framework of the ITER-RFX Neutral Beam Testing Facility (NBTF) Agreement and has received funding from the ITER Organization. The views and opinions expressed herein do not necessarily reflect those of the ITER Organization.

This work has also been carried out within the framework of the EUROfusion Consortium, funded by the European Union via the Euratom Research and Training Programme (Grant Agreement No 101052200 — EUROfusion). Views and opinions expressed are however those of the author(s) only and do not necessarily reflect those of the European Union or the European Commission. Neither the European Union nor the European Commission can be held responsible for them.

**Corresponding author: riccardo.casagrande@igi.cnr.it**